\documentclass[sigconf]{acmart}
\AtBeginDocument{%
  \providecommand\BibTeX{{%
    \normalfont B\kern-0.5em{\scshape i\kern-0.25em b}\kern-0.8em\TeX}}}

\copyrightyear{2020}
\acmYear{2020}
\setcopyright{acmlicensed}\acmConference[KDD '20]{Proceedings of the 26th ACM SIGKDD Conference on Knowledge Discovery and Data Mining}{August 23--27, 2020}{Virtual Event, CA, USA}
\acmBooktitle{Proceedings of the 26th ACM SIGKDD Conference on Knowledge Discovery and Data Mining (KDD '20), August 23--27, 2020, Virtual Event, CA, USA}
\acmPrice{15.00}
\acmDOI{10.1145/3394486.3403199}
\acmISBN{978-1-4503-7998-4/20/08}

\usepackage{booktabs} 
\usepackage{graphicx} 
\usepackage{subcaption}
\usepackage[utf8]{inputenc} 
\usepackage[T1]{fontenc}    
\usepackage{hyperref}       
\usepackage{url}            
\usepackage{booktabs}       
\usepackage{amsfonts}       
\usepackage{nicefrac}       
\usepackage{microtype}      
\usepackage{amsthm}
\usepackage{algorithm,algpseudocode}
\RequirePackage{amsmath, subcaption}
\RequirePackage{amsmath,dsfont}
\usepackage{amsthm}
\usepackage{amssymb,accents}
\usepackage{graphicx}
\usepackage{amssymb}
\usepackage{ulem}
\usepackage{graphicx}
\usepackage{geometry}
\usepackage{microtype}
\usepackage{graphicx}
\usepackage{booktabs} 
\usepackage[font=small,labelfont=bf]{caption}
\usepackage{algorithm,algpseudocode}
\usepackage{multirow}

\usepackage{caption}
\usepackage{caption}
\newtheorem{theorem}{Theorem}

\input{Definitions}

\begin{document}
\fancyhead{}

\title{Evaluating Fairness Using Permutation Tests}

\author{Cyrus DiCiccio}
\affiliation{%
  \institution{LinkedIn Corporation}
}
\email{cdiciccio@linkedin.com}

\author{Sriram Vasudevan}
\affiliation{%
  \institution{LinkedIn Corporation}
}
\email{svasudevan@linkedin.com}

\author{Kinjal Basu}
\affiliation{%
  \institution{LinkedIn Corporation}
}
\email{kbasu@linkedin.com}

\author{Krishnaram Kenthapadi$^{1}$}\thanks{$^1$Work done while at LinkedIn}
\affiliation{%
  \institution{Amazon AWS AI}
}
\email{kenthk@amazon.com}

\author{Deepak Agarwal}
\affiliation{%
  \institution{LinkedIn Corporation}
}
\email{dagarwal@linkedin.com}
\renewcommand{\shortauthors}{DiCiccio, et al.}

\begin{abstract}
	
Machine learning models are central to people's lives and impact society in ways as fundamental as determining how people access information. The gravity of these models imparts a responsibility to model developers to ensure that they are treating users in a fair and equitable manner. Before deploying a model into production, it is crucial to examine the extent to which its predictions demonstrate biases. This paper deals with the detection of bias exhibited by a machine learning model through statistical hypothesis testing.  We propose a permutation testing methodology that performs a hypothesis test that a model is fair across two groups with respect to any given metric. There are increasingly many notions of fairness that can speak to different aspects of model fairness. Our aim is to provide a flexible framework that empowers practitioners to identify significant biases in any metric they wish to study. We provide a formal testing mechanism as well as extensive experiments to show how this method works in practice.

\end{abstract}


\maketitle

\section{Introduction}
\label{sec:intro}
Machine learned models are increasingly being used in web applications for crucial decision-making tasks such as lending, hiring, and college admissions, driven by a confluence of factors such as ubiquitous connectivity, the ability to collect, aggregate, and process large amounts of fine-grained data, and the ease with which sophisticated machine learning models can be applied. Recently, there has been a growing awareness about the ethical and legal challenges posed by the use of such data-driven systems, which often make use of classification models that deal with users. Researchers and practitioners from different disciplines have highlighted the potential for such systems to discriminate against certain population groups, due to biases in data and algorithmic decision-making systems. Several studies have shown that classification and ranked results produced by a biased machine learning model can result in systemic discrimination and reduced visibility for an already disadvantaged group~\cite{nips_2017_tutorial, dwork_2012, hajian_2016_tutorial, pedreschi_2009} (e.g., disproportionate association of higher risk scores of recidivism with minorities~\cite{angwin_2016}, over/under-representation and racial/gender stereotypes in image search results~\cite{kay_2015}, and incorporation of gender and other human biases as part of algorithmic tools~\cite{bolukbasi_2016, caliskan_2017}). One possible reason is that machine-learned prediction models that are trained on datasets exhibiting existing societal biases end up learning these biases, and can therefore reinforce (or even potentially amplify) them in its results.

Our goal is to develop a framework for identifying biases in machine-learned models across different subgroups of users, and address the following questions:
\begin{itemize}
\item Is the measured discrepancy statistically significant? 
When dealing with web-scale datasets, we are very likely to observe discrepancies of varying magnitudes owing to less-than-ideal scenarios and noise. Observed discrepancies do not necessarily imply that there is bias - the strength of the evidence (as presented by the data) must be considered in order to ascertain that there truly is bias. To this end, we seek to perform rigorous statistical hypothesis tests to quantify the likelihood of the observed discrepancy being due to chance.  

\item Can we perform hypothesis tests in a metric-agnostic manner? When certain assumptions about the underlying distribution or the metric to be measured can be made, we can resort to parametric tests suited for these purposes. However, when we wish to have a pluggable interface for any metric (with respect to which we wish to measure discrepancies in fairness), we need to make the testing framework as generic as possible.
\end{itemize}

There are numerous definitions of fairness including equalized odds, equality of opportunity, individual or group fairness, and counterfactual fairness in addition to simply comparing model assessment metrics across groups.  While each of these criteria has merit, there is no consensus on what qualifies a model as fair, and this question is beyond the scope of this paper.  Our aim is not to address the relative virtues of these definitions of fairness, but rather to assess the strength of the evidence presented by a dataset that a model is unfair with respect to a given metric.   

We develop a permutation testing framework that serves as a black-box approach to assessing whether a model is fair with respect to a given metric, and provide an algorithm that a practitioner can use to quantify the evidence against the assumption that a model is fair with respect to a specified metric. 
This is especially appealing because the framework is metric agnostic. 

Traditional permutation tests specify that the underlying data-generating mechanisms are identical between two populations and are somewhat limited in the claims that can be made regarding the fairness of machine learning models. We seek to determine whether a machine learning model has equitable performance for two populations in spite of potential inherent differences between these populations.  In this paper, we illustrate the shortcomings of classical permutation tests, and propose an algorithm for permutation testing based on any metric of interest which is appropriate for assessing fairness.  Open source packages evaluating fairness (such as \cite{tramer2017}) implement permutation tests which are not valid for their stated use.  Our contribution is to illustrate the potential pitfalls in implementing permutation tests and to develop a permutation testing methodology which is valid in this context.

The rest of the paper is organized as follows. In Section \ref{sec:background}, we provide a background on permutation tests and illustrate why traditional permutation tests can be problematic as well as how to solve these issues. Section \ref{sec:methodology} introduces permutation tests that can evaluate fairness in machine learning models. Simulations are presented in Section \ref{sec:simulation}, followed by experiments using real-world datasets in Section \ref{sec:experiment}. We discuss related work in Section \ref{sec:related} and conclude in Section \ref{sec:conclusion}. The proofs of all results are pushed to the Appendix for ease of readability.

\section{Preliminaries}
\label{sec:background}
Permutation tests (discussed extensively in \cite{good2000permutation}) are a natural choice for making comparisons between two populations; however, the validity of permutation tests is largely dependent on the hypothesis of interest, and these tests are frequently misapplied.  We describe some background, illustrate misapplications of permutation tests, and establish valid permutation tests in the context of assessing the fairness of machine learning models.  

\subsection{Background 
}

The standard setup of two sample permutation tests is as follows.  A sample $Y_1,...,Y_{n_Y}$ is drawn from a population with distribution $P_Y$ and a sample $X_1,...,X_{n_X}$ is drawn from a population with distribution $P_X$.  The null hypothesis of interest is
$$H_0:P_X=P_Y$$
which is often referred to as either the ``strong'' or ``sharp'' null hypothesis.  A common example is comparing two drugs, perhaps a treatment with a placebo to study the effectiveness of a new drug.  The observed data is some measure of outcome for each group given either the treatment or control.  In this case, the null hypothesis is that there is no difference whatsoever in the observed outcomes between the two groups.  

A test statistic $T(X_1,..., X_{n_X}, Y_1,..., Y_{n_Y})$ (such as $T(X, Y) =\bar X-\bar Y$) is chosen based on the effects that the researcher would like to detect, and the distribution of this statistic under the null hypothesis is approximated by repeatedly computing the statistic on permuted samples as follows:
\begin{itemize}
	\item For a large integer $B$, uniformly choose $B$ permutations $\pi_1,...,\pi_B$ of the integers $\left\{1,...,n_X+n_Y\right\}$
	\item Define $(Z_1,...,Z_{n_X+n_Y}) = (X_1,...,X_{n_X}, Y_1,...,Y_{n_Y})$.
	\item Recompute the test statistic on the permutations of the data resulting in $T_{\pi_i} = T(Z_{\pi_i(1)},...,Z_{\pi_i(n_X+n_Y)})$.
	\item Define the permutation distribution of $T$ to be the empirical distribution of the test statistics computed on the permuted data, namely $$\hat P (T \leq t) = \frac{1}{B}\sum_{i=1}^B I \left\{ T_{\pi_i} \leq t \right\}.$$
	\item Reject the null hypothesis at level $\alpha$ if $T$ exceeds the $1-\alpha$ quantile of the permutation distribution.
\end{itemize}

This test is appealing because it has an exactness property: when the ``sharp'' null hypothesis is true, the probability that the test rejects is exactly $\alpha$ (Type I error rate).  However, researchers are more commonly interested in testing a ``weak'' null hypothesis of the form
$$H_0:\theta(P_X)=\theta (P_Y)$$
where $\theta(\cdot)$ is some functional, parameter, etc. of the distribution.  Furthermore, researchers typically desire assigning directional effects (such as concluding that $\theta(P_X)>\theta(P_Y)$) in addition to simply rejecting the null hypothesis. For instance, in the case of comparing two drugs, the null hypothesis may specify that the mean recovery times are identical between the two drugs:
$H:\mu_X=\mu_Y$.
In the case of rejecting, the researcher would like to conclude either $\mu_X>\mu_Y$ or $\mu_X<\mu_Y$ so that recommendations for the more efficacious drug can be given. Merely knowing that there is a difference between the drugs but being unable to conclude which one is better would be unsatisfying.  

While the permutation test is exact for the strong null hypothesis, this is not the case for the weak null.  Depending on the test statistic used, the permutation test may not be valid (even asymptotically) for the weak null hypothesis: the rejection probability can be arbitrarily large when only the weak null hypothesis is true (larger than the specified level, as is the requirement for a valid statistical test). This leads to a much higher Type I error rate than expected.

\subsection{An Illustrative Example
}

We use a simple example of comparing means to illustrate the problems with permutation tests. 
Suppose that $X_1,...,X_{n_X} \sim N(\mu_X, \sigma^2_X)$ and $Y_1,...,Y_{n_Y} \sim N(\mu_Y, \sigma^2_Y)$ with $n_X/n_Y = \lambda$.  Suppose the test statistic used is $T(X,Y) = \sqrt{n_X} (\bar X- \bar Y)$.  Note that the scaling is chosen to have a non-degenerate limiting distribution.  The sampling distribution of $T$ is asymptotically normal with mean 0 and variance $n_X\left(\text{var}\left(\bar X\right) + \text{var}\left(\bar Y\right)\right) = \sigma_X^2+\lambda \sigma_Y^2$.  

When permuting the data, samples from both populations $X$ and $Y$ are taken (without replacement) from the pooled distribution, and the permutation distribution behaves as though both samples were taken from a mixture distribution
$$p P_X+(1-p)P_Y \sim p N(\mu_X, \sigma^2_X)+(1-p)N(\mu_Y, \sigma^2_Y)$$
where $p = n_X/(n_X+n_Y)$.  The variance of this mixture distribution is 
$$\sigma^2_{pooled} =p\sigma_X^2+(1-p)\sigma_Y^2 $$
Thus, for our chosen statistic, the permutation distribution is (asymptotically) normal with mean 0 and variance
$$\sigma^2_{pooled}+\lambda \sigma^2_{pooled}.$$
Under the strong null (specifying equality of distributions), $\sigma^2_X=\sigma^2_Y=\sigma^2_{pooled}$ and the permutation distribution approximates the sampling distribution.  However, under the weak null, it may be the case that $\sigma^2_X \neq \sigma^2_Y$ and consequently $\sigma^2_X \neq \sigma^2_Y \neq \sigma^2_{pooled}$. Under the weak null, this means that the permutation distribution cannot be used as an approximation of the sampling distribution, and any inference based on the permutation distribution is therefore invalid.   

\subsection{Valid Permutation Tests}
Choosing a pivotal (asymptotically distribution-free; does not depend on the observed data's distribution) statistic can rectify the issue as identified above.  For instance, the sampling distribution of the Studentized statistic 
$$T(X,Y)=\frac{\sqrt{n_x}(\bar X - \bar Y)}{\sqrt{s_X^2+s_Y^2 n_X/n_Y}},$$
where $s_X$ and $s_Y$ are the sample standard deviations, is asymptotically $N(0,1)$. The statistic is pivotal because the asymptotic distribution does not rely on the distributions of the observed data. Since it is distribution-free, the permutation distribution of the Studentized statistic (which behaves as though the two groups were sampled from a distribution that is not necessarily the same as the underlying distributions of these two groups) is asymptotically standard normal as well.
Typically, Studentizing the test statistic will give validity for testing with an asymptotically normal test statistic. If the choice of the test statistic is asymptotically pivotal, the resulting permutation test can be expected to be asymptotically valid.

Note that even if we are interested in testing the strong null hypothesis, but wish to make directional conclusions, the directional errors can be quite large with the un-Studentized statistic.  This can occur, for instance, when there is a small positive effect, and the difference of means is negative, but the test rejects due to a difference in standard deviations.  The Studentized statistic ensures that the chance of a directional error is no larger than $\alpha/2$.

\section{Permutation Tests for Fairness}
\label{sec:methodology}
Based on the discussion above, let us now consider the problem of testing fairness of a machine learning model on two groups, say group A and group B.  We may want to compare metrics such as the area under the receiver operator curve (AUC) between the two groups.  In this setting, the permutation test is exact for testing the strong null hypothesis that the distribution of the observed data (the features used by the model together with the outcome) arise from the same distribution between the two groups.

More concretely, suppose that the generated data for group A is $(X_{A,1},y_{A,1}),...,(X_{A,n_A},y_{A,n_A})$ and the data for group B is $(X_{B,1},y_{B,1}),$ $...,(X_{B,n_B},y_{B,n_B})$ where the $X$'s are $p$-dimensional vectors of available features and the $y$'s are response variables. The strong null hypothesis specifies
$$H_0:P_{(X_A,y_A)} = P_{(X_B,y_B)} .$$
That is, the joint distributions of the features and the response are identical between the two groups.  Surely if this holds, the model will be fair, but a permutation test may reject (or fail to) based on differences in the distribution of features rather than inherent fairness of the models.
To further illustrate the issues of permutation testing for fairness, we will discuss tests of fairness based on several statistics for binary classifiers before presenting an algorithm for testing fairness for a general classification or regression problem. 

\subsection{Within Outcome Permutation}
In the context of comparing binary classifiers, one can perform permutation tests under weaker assumptions by permuting data only within positive or negative labels.  
The distribution of binary classifier metrics typically depends on the number of observations in the positive and negative categories.  A slightly more generally applicable permutation test can be performed by permuting the positive examples between groups A and B and separately permuting the negative examples between both groups.  This is valid under the slightly more general assumption that the distributions of the features corresponding to the positive examples are equal between the two groups, and likewise for the negative examples. This method of permuting the data is valid under the null hypothesis 
$$H:P_{X_A^+}=P_{Y_A^+} \text{ and } H:P_{X_A^-}=P_{Y_A^-}$$  
i.e. under the null hypothesis that the distribution of the features for the positive (and negative) labeled data are equal.  This is slightly more flexible than merely permuting labels in that it does not require that the proportion of positive and negative labels be the same between the groups, but still retains exactness.  

\subsection{Comparing AUC}
In the context of testing for equality of AUC, the null hypothesis of interest is that the AUC of the model is equal between the two groups:
$$H:AUC_A=AUC_B.$$
For notational convenience, we use $X_{A,1}^+, \ldots, X_{A,n_A^+}^+$ and $X_{A,1}^-$, $\ldots$, $X_{A,n_A^-}^-$ for the observed features of the $A$ group corresponding to positive and negative labels respectively, with $n_A^+ + n_A^-=n_A$ (and similarly for group B).  
Assuming that the classifier of interest assigns a positive outcome if some function $f(\cdot)$ exceeds a threshold, the null hypothesis of equality of AUC's is equivalent to testing
$$H:P(f(X_{A,1}^+) > f(X_{A,1}^-)) = P(f(X_{B,1}^+) > f(X_{B,1}^-)).$$
This can be done by using the Mann-Whitney (MW) statistic,
\begin{align*}
T_{MW} = & \sqrt{n} \left( \frac{1}{n^+_An^-_A} \sum_{i,j} I \left\{ f(X_{A,i}^+) > f(X_{A,j}^-) \right\} \right. \\
&\qquad - \left. \frac{1}{n^+_Bn^-_B} \sum_{i,j} I \left\{ f(X_{B,i}^+) > f(X_{B,j}^-) \right\}\right).
\end{align*}
When using this statistic to perform a permutation test, the permutation distribution behaves as though both samples are taken from the mixture distribution
$$p_A P_{(X_A,y_A)}+(1-p_A)P_{(X_B,y_B)}$$
where $p_A = n_A/(n_A+n_B)$.  In general, the permutation distribution of the $T_{MW}$ under this mixture distribution will not be equal to the sampling distribution when only the assumption of equality of AUC’s holds.  The variance of the sampling distribution (as stated in the next section) is dependent on the proportion of positive and negative examples, so if the two groups differ inherently in the number of positive and negative examples, the test will not be valid.  Furthermore, using the permutation test will not be valid for making directional conclusions, which in this case is determining which group the model is biased towards.

\subsubsection{Validity of the Studentized Difference of AUCs}
The MW test statistic has variance $\VV_A(P_{(X_A,y_A)})/p_A+\VV_B(P_{(X_B,y_B)})/(1-p_A)$ where
\begin{align*}
\VV_A(P_{(X_A,y_A)})= & \frac{1}{1-p_{+,A}}P(f(X_{A,1}^+) > f(X_{A,1}^-), f(X_{A,1}^+) > f(X_{A,2}^-)) \\
& + \frac{1}{p_{+,A}} P(f(X_{A,1}^+) > f(X_{A,1}^-),f(X_{A,2}^+) > f(X_{A,1}^-)) \\
&  - \left( \frac{1}{p_{+,A}} +  \frac{1}{1-p_{+,A}}\right) P(f(X_{A,1}^+) > f(X_{A,1}^-))^2,\\
\end{align*}
$X_{A,1}^\pm$ and $X_{A,1}^\pm$ are independently randomly selected feature vectors, $p_{+,A} = \EE \left[I \left\{ y_{i,A} = 1 \right\}\right]$, and $\VV_B$ is defined analogously.  The asymptotic normality of the statistic is given by the following Theorem

\begin{theorem}\label{thm:AUC}
Let $P^{MW}_{n_A, n_B}$ denote the sampling distribution of the Mann-Whitney statistic under the null hypothesis.  Then 
\[
\sup_{t \in \mathbb{R}} |P^{MW}_{n_A, n_B}(t) - \Phi(t/\sqrt{v})| \rightarrow 0
\]
where $v = \VV_A(P_{(X_A,y_A)})/p_A +\VV_B(P_{(X_B,y_B)})/(1-p_A)$.  Let $\hat P^{MW}_{n_A, n_B}$
denote the permutation distribution of the MW statistic.  Then 
\[
\sup_{t \in \mathbb{R}} |\hat P^{MW}_{n_A, n_B}(t) - \Phi(t/\sqrt{w})| \rightarrow 0
\]
in probability, where $w = \VV_A(p_A P_{(X_A,y_A)}+ (1-p_A)P_{(X_B,y_B)} )/p_A +\VV_B(p_A P_{(X_A,y_A)} + (1-p_A)P_{(X_B,y_B))}/(1-p_A)$.
 \end{theorem}

The permutation distribution behaves as though each sample was taken from the mixture distribution $p_A P_{(X_A,y_A)} + (1-P_A)P_{(X_B,y_B)}$ and may not approximate the sampling distribution.  In particular, the variance of the permutation distribution is not necessarily the same as that of the sampling distribution. 

These variances can be consistently estimated, for example by using DeLong's method.  The sampling distribution of the ``Studentized'' MW test statistic, which normalized by a consistent estimator of the variance, is asymptotically standard normal.  Using the Studentized test, the permutation distribution is asymptotically standard normal.  

\begin{theorem}\label{thm:AUCstudentized}
	Let $Q^{MW}_{n_A, n_B}$ and $\hat Q^{MW}_{n_A, n_B}$ denote the sampling distribution and permutation distribution, respectively of the Mann-Whitney statistic normalized by a consistent estimate of the sample standard deviation.  Then 
	\[
	\sup_{t \in \mathbb{R}} |\hat Q^{MW}_{n_A, n_B}(t) - Q^{MW}_{n_A, n_B}(t)| \rightarrow 0
	\] 
	in probability.
\end{theorem}

Because the Studentized statistic is asymptotically pivotal, the permutation distribution and the sampling distribution have the same limiting behavior, and hence the permutation distribution approximates the sampling distribution.  

\subsection{Proportion Statistics}\label{sec:proportion-stats}

Although permutation tests are typically valid for comparing proportions, permutation tests may be problematic for binary classification metrics which appear to be measuring proportions (such as true or false positive rate, sensitivity, specificity, etc.).  A simple illustrative, classical example is comparing two Bernoulli proportions between independent samples.  In this case, the proportion uniquely specifies the distribution of the random variables, and the null hypothesis of equality of distributions is equivalent to testing equality of proportions.   In this case, the usual permutation test is valid.  

Binary classification metrics such as false positive (or negative) rate, sensitivity, specificity, etc. which seems at face value to be a straight-forward extension of the two proportion z-test, can be problematic.  We will focus our discussion on false-negative rate, however, obvious extensions hold for other metrics.

Suppose that the decision of the classifier is given by a function $c(\cdot)$ (so that $c(x)=1$ if an observation with features $x$ is classified as positive example and $c(x)=-1$ if the observation is classified as a negative example).  Define
$$\hat p_A^-=\frac{1}{n_A^+}\sum_{i=1}^{n_A^+} I \left\{c(X_{i,A}^+) =-1\right\}$$
to be the (empirical) false negative rate for group A, and define $\hat p_B^-$ analogously for group B.
In the above notation, the difference of false negative rates is
$\hat p_A^- - \hat p_B^-$. Scaled by say $\sqrt{n}$ and assuming that for each group $$\EE I\left\{c(X_{i,g}^+) =-1\right\} = p_{FN,g},$$ the asymptotic distribution of the statistic $T(X,Y) = \sqrt{n}\left(\hat p_A^- - \hat p_B^-\right)$ is
$$N\left( 0, \frac{p_{FN,A}(1-p_{FN,A})}{p_A p_{+,A}} + \frac{p_{FN,B}(1-p_{FN,B})}{(1-p_A) p_{+,B}} \right)$$
where $p_{+,A} = \EE I \left\{ y_{i,A} = 1 \right\}$ and $p_A = n_A/n$ with $n=n_A+n_B$.  

When permuting labels of the data, the number of positive examples assigned to group A or group B will differ for each permutation.  Heuristically, the permutation distribution behaves as though each group was sampled from the mixture distribution between the two groups.  In particular, the permutation distribution approximates the sampling distribution when the proportion of positive examples is the same between the two groups.  The permutation distribution is asymptotically normal with mean $0$ and variance 
$$\frac{p_{FN,A}(1-p_{FN,A})}{p_A p_{+}} + \frac{p_{FN,B}(1-p_{FN,B})}{(1-p_A) p_{+}} $$
where $p_+ = p_A \cdot p_{+,A} + (1-p_A) \cdot p_{+,B}$.  In general, this variance will not be equal to the asymptotic variance of the sampling distribution and the permutation test may fail to be valid.

A general fix for permutation tests is to again use an asymptotically pivotal test statistic (one whose asymptotic distribution does not depend on the distributions generating the two groups).  In this example, using a Studentized test statistic:
$$S(X,Y) = \frac{\sqrt{n}(\hat p_A^-- \hat p_B^-)}{\sqrt{\frac{\hat p_{FN,A}(1-\hat p_{FN,A})}{p_A \hat p_{+,A}} + \frac{\hat p_{FN,B}(1-\hat p_{FN,B})}{(1-p_A) \hat p_{+,B}} }}$$
(where hats denote the usual sample proportions) will cure the problem.  When using this Studentized statistic, both the permutation distribution and sampling distribution are asymptotically standard normal which ensures the test is asymptotically valid.  
\begin{theorem}\label{thm:FNR}
	Assume that for both groups the probability of observing a positive example is non-zero, and the probability of correctly classifying these examples is bounded away from 0 and 1.  Denote by $\hat P_{n_A, n_B}(\cdot)$ and $P_{n_A, n_B}(\cdot)$ the permutation distribution of the Studentized statistic $S(X,Y)$ and the sampling distribution under the null hypothesis, respectively, for a sample of size $n_A$ from group A and $n_B$ from group B.  Then, 
	$$\sup_{t \in \mathbb{R}} \left| \hat P_{n_A, n_B}(t) - P_{n_A, n_B}(t) \right| \rightarrow 0$$
	almost surely.
\end{theorem}

While this result pertains to false-negative rate, permutation tests based on Studentized statistics are generally valid.  Verifying the validity of other proportion-based binary classification metrics is similar to the example of the false-negative rate.  

\subsection{A General Algorithm}
The examples presented above demonstrate the need for studentizing statistics in binary classification problems.  We now provide a general algorithm for performing permutation tests of fairness that is applicable to general classification problems, as well as regression problems.  

In the examples presented above, closed form, consistent variance estimators of the test statistic are easily computed.  In examples where finding a covariance estimator is difficult, a bootstrap variance estimator can be used. Supposed that data $D_A$ is observed from group A and $D_B$ is observed from group B. For a fixed a large integer $n_b$, if we resample with replacement $n_A$ samples $D^*_{A, i}$ from $D_A$ and $n_B$ samples $D^*_{B, i}$ from $D_B$ for $i=1,...,n_b$, then the bootstrap estimate of the variance of a statistic $T$ is 
\begin{align}
\label{eq:bootstrap_var}
\VV(D_A, D_B) = \frac{1}{n_b}\sum_{i =1}^{n_b} \left( T(D^*_{A,i}, D^*_{B,i}) -  T(D_A, D_B) \right)^2
\end{align}
where $n_b$ denotes the number of bootstrap trials. Whether the test statistic is asymptotically pivotal or needs to be Studentized, a valid permutation test can be performed according to the following algorithm.

\begin{algorithm}[!h]
	\caption{Evaluating Fairness via Permutation Testing}\label{algo:largescale}
	\begin{algorithmic}[1]
		\State \text{Input: } Test statistic $T$, data $D_A$ from group A and $D_B$ from group B, number of permutation samples $n_p$  
		\State Output : Permutation p-value $\hat p$
		\For{$i=1,...,n_p$}
		\State Choose a random permutation $\pi_i$
		\State Compute the test statistic $T_i = T(D_{A,\pi_i}, D_{A,\pi_i})$
		\If{$T$ is not asymptotically pivotal}  
		\State Estimate the variance $v_i = \VV(D_{A,\pi_i}, D_{A,\pi_i})$ 
		 as in \eqref{eq:bootstrap_var}
		\State Set $S_i = T_i/\sqrt{v_i}$ 
		\EndIf
		\EndFor
		\If{$T$ is asymptotically pivotal}  
		\State Compute $\hat p = \sum_{i=1}^{n_p} I \left\{ T(D_A,D_B) > T_i \right\}$
		\Else
		\State Estimate the variance of $T$ ($\VV(D_A, D_B)$) as in \eqref{eq:bootstrap_var}
		\State Set $S(D_A, D_B) = T(D_A, D_B)/\sqrt{\VV(D_A, D_B)}$
		\State Compute $\hat p = \sum_{i=1}^{n_p} I \left\{ S(D_A,D_B) > S_i \right\}$
		\EndIf
		\State
		\Return $\hat p$, a p-value for evaluating fairness with respect to the 
		provided metric.
	\end{algorithmic}
\end{algorithm}

\section{Simulation Study}
\label{sec:simulation}
Most papers on Fairness in Machine Learning focus on a single definition of fairness, and many adjust model training in order to reduce unfairness\cite{agarwal2018reductions, mary2019fairness}. We could not find any literature on statistically testing if a model is fair other than FairTest \cite{tramer2017}. Through simulations, we compare our methodology to FairTest and illustrate some of the problems we are able to cure. Moreover, we show detailed simulations to demonstrate the need for Studentizing the test statistic and compare it to the bootstrap method.

\subsection{Comparison With FairTest}
FairTest \cite{tramer2017} uses a permutation test based on Pearson's correlation statistic.  
We demonstrate the issues of the permutation test as implemented in FairTest (which is based on an un-Studentized test statistic), both for testing correlation (which is the stated use) and independence between a protected attribute and model error.  The permutation test implemented by FairTest is neither valid for testing the correlation between a protected attribute and model error 
nor very powerful for testing independence. 
To demonstrate these issues, we must know the ground truth of the model we are using, so we prefer to use simulated data rather than experimental data. 

We first provide an example demonstrating that FairTest's implementation is not a valid test for their stated use, but that Algorithm \ref{algo:largescale} provides a valid test. Suppose the protected attribute, $x$ is generated as a uniform random variable (bounded away from zero to avoid dividing by values near 0)
\[
x \sim U(10^{-5},1)
\]
and the prediction error for a model is given as 
\begin{align}
\label{eq:errordist}
\epsilon = \frac{\tilde \epsilon}{x^2} \text{ where } \tilde \epsilon \sim N(0, 1)
\end{align}
such a setting may be a reasonable approximation to $x$ being a normalized age variable.  In this model, the protected attribute and model error are uncorrelated, although the model error depends heavily on the protected attribute, so we do not have independence.  Therefore, the rejection probability for the test of uncorrelatedness at say the 5\% nominal level (i.e. a test at a 5\% level of significance), should have a null rejection probability of $0.05$ in this setting since the null hypothesis is indeed true.  Table \ref{tab:rejProb} reports a Monte-Carlo approximation to the null rejection probability by generating 2000 samples of the protected attribute $x$ and model errors, and performing a permutation test using FairTest's implementation (based on Pearson's correlation) and our methodology (from Algorithm \ref{algo:largescale} based on the studentized Pearson's correlation).  The reported probability of rejecting the null hypothesis is based on 1,000 permutations of the data and averages the test decisions over 10,000 such simulations. 


\begin{table}[ht]
\centering
\begin{tabular}{|c|c|c|c|}
\hline
Null & \multicolumn{3}{c|}{Rejection Probability} \\
Hypothesis & Algorithm \ref{algo:largescale} & FairTest & Desired\\
\hline 
Uncorrelated & 0.0428 & 0.7508 & 0.05\\
Independence & 0.0501 & 0.0099 & $\geq$ 0.05\\
\hline
\end{tabular}
\caption{Comparison of Null Rejection Probabilities}
\label{tab:rejProb}
\end{table}

In this example of testing for uncorrelatedness, the rejection probability of either testing methodology should be equal to (or at least below) the nominal level (0.05) since the null hypothesis is indeed true.  We find the null rejection probability for FairTest is dramatically above the nominal level, reaffirming that the test is not valid for testing uncorrelatedness, whereas the test using Algorithm \ref{algo:largescale} has a null rejection probability close to the nominal level.   

Even if the practitioner desires to test the null hypothesis of independence between the protected attribute and model error, the test implemented by FairTest is ``biased,'' a statistical term meaning that the rejection probability can be below the nominal level of the test for some alternatives.  For instance, if the test is performed at the 5\% level, the rejection probability can be dramatically below 0.05 when the null hypothesis of independence is not true.  Practically, this means that the power to detect unfairness in a dataset can be significantly worse than random guessing.  It is also important to note that even in web-scale settings, where the issue is often determining practical significance rather than statistical significance, the naive permutation may fail to detect bias because of the lack of power illustrated here.  

To give a concrete example, suppose that the protected attribute is generated as an exponential distribution with rate parameter one (plus one to avoid the instability of dividing by values near zero),
\[
x \sim Exp(1)+1
\]
and the prediction error for a model is again given as \eqref{eq:errordist}.

Performing a Monte-Carlo simulation analogous to the setting testing for uncorrelatedness, the null rejection probability of the test for independence is given in Table \ref{tab:rejProb}.  As we can see, the rejection probability for FairTest is substantially below the nominal level 5\% (despite the fact that the attribute and error are dependent), whereas that of the p-value using Algorithm \ref{algo:largescale} is that of the nominal level.  While the test using Algorithm \ref{algo:largescale} is not powerful in this setting, it is at least unbiased which is a first-order requirement for a reasonable statistical test.  The lack of power comes from the choice of test statistic which can easily be changed if the practitioner desires power to detect dependence among correlated variables.

Depending on the hypothesis of interest, the permutation testing implementation given in FairTest can be either invalid or biased.  On the other hand, Algorithm \ref{algo:largescale} provides a test that is valid for correlation and unbiased for independence.

\begin{figure*}[ht]
\centering
\begin{subfigure}{.3\textwidth}
  \centering
  \includegraphics[width= 0.9\linewidth]{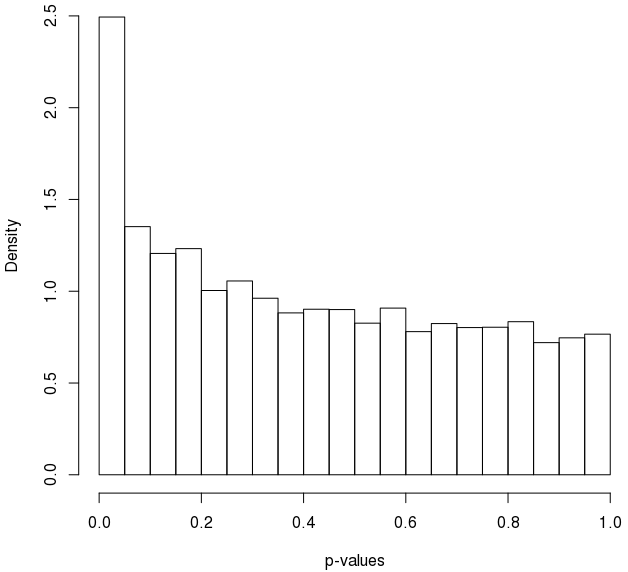}
  \caption{Unstudentized Test Statistic}
  \label{fig:fnr2}
\end{subfigure}%
\begin{subfigure}{.3\textwidth}
  \centering
  \includegraphics[width= 0.9\linewidth]{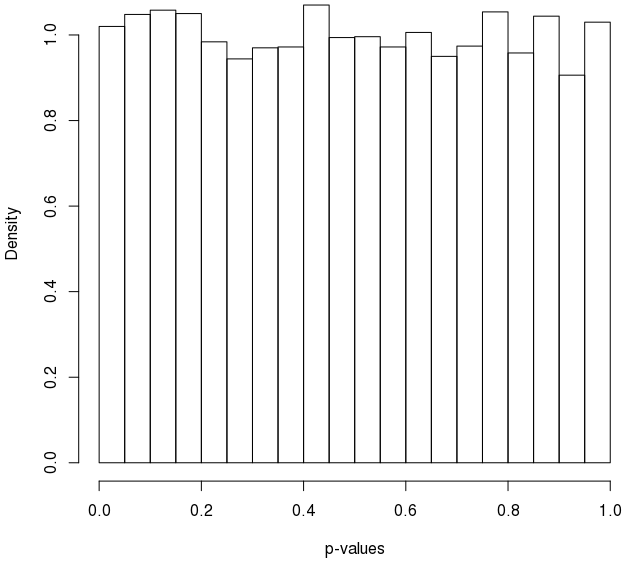}
  \caption{Studentized Test Statistic}
  \label{fig:fnrstud2}
\end{subfigure}
\begin{subfigure}{.3\textwidth}
  \centering
 	\includegraphics[width= 0.9\linewidth]{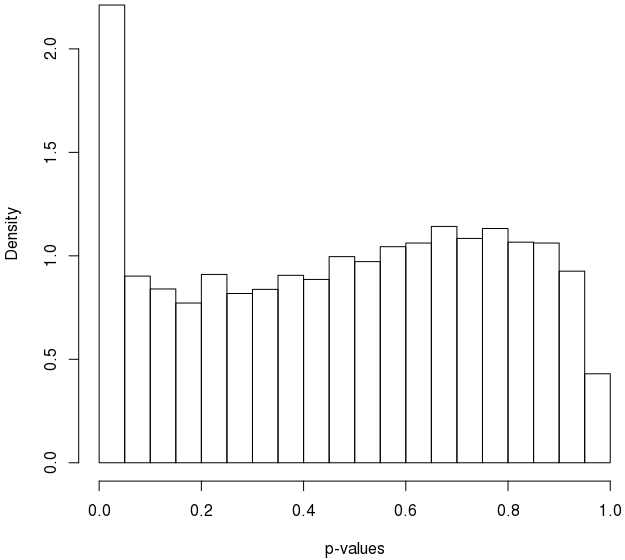}
	\caption{Bootstrap $p$-values for difference of FNRs}
	\label{fig:fnrbs}
\end{subfigure}
\caption{Histogram of $p$-values for difference of FNRs via different statistics}
\label{fig:hist}
\end{figure*}

\subsection{Comparison of Permutation Methods And The Bootstrap}

We consider a case of comparing the false-negative rate of group A with group B.  For group A, $n_A = 200$ samples are generated, with an $80\%$ chance of observing a positive outcome and a $20\%$ chance of observing a negative outcome.  For group B, $n_B= 200$ samples are generated, with a $20\%$ chance of observing a positive outcome and an $80\%$ chance of observing a negative outcome.  For both groups, the classifier has a true positive (and true negative) rate of $90\%$.  The classifier is fair in the sense that the false-negative rate is equal between the two groups.  For $10,000$ simulations, data is generated in this manner and a permutation p-value is obtained using both the studentized and un-studentized statistics.  Figures \ref{fig:fnr2} and \ref{fig:fnrstud2} give histograms of the p-values using the un-studentized and studentized statistic, respectively.  Note that the p-values should be uniform, and the p-values using the Studentized statistic are much closer to the uniform distribution.  At nominal level $\alpha = 0.05$, the rejection probability using the un-studentized statistic is $0.1216$ (very anti-conservative) and the rejection probability using the studentized statistic is $0.0486$ (very nearly exact). 

%
%
%
Keeping in mind the goal of providing a metric agnostic system for inference regarding fairness, another natural choice of methodology would be to implement a bootstrap.  In this case, we compare our approach with the ``basic'' bootstrap, implemented as follows:

\begin{itemize}
	\item Uniformly resample, with replacement, $(X_{A,1}^*, y_{A,1}^*)$,
	...,\\$(X_{A,n_A}^*, y_{n_A,A}^*)$ from $(X_{A,1}, y_{A,1}),...,(X_{A,n_A}, y_{n_A,A})$ and similarly take a uniform sample, with replacement, from group B.  
	\item If $T$ is a test statistic of interest, approximate the distribution of $T$ using the distribution of $T^*-T$ where $T^*$ is computed on the resampled data.  
\end{itemize}

The basic bootstrap has the advantage that the statistic need not be studentized; however, it has no guarantees of exactness.  In the simulation setting described above, the null rejection probability using the un-studentized difference of false-negative rates is $0.1078$.  The distributional approximation using the bootstrap is considerably worse than the permutation test based on the studentized statistic, so we recommend using a permutation test over a bootstrap (see Figure \ref{fig:fnrbs}). 
\section{Real-World Experiments}
\label{sec:experiment}
The permutation testing framework as described above was implemented in Scala, to work with machine learning pipelines that make use of Apache Spark \cite{apacheSpark}. The framework supports plugging in arbitrary metrics or statistics whose difference is to be compared, such as precision, recall or AUC. To studentize the observed difference of the statistic between the two groups, we need to estimate its standard deviation. We achieve this by performing a bootstrap to obtain the distribution of these differences and computing an unbiased estimate of the variance, from which we obtain the standard deviation.

To studentize the differences obtained during the permutation trials, we make use of the standard deviation of the permutation distribution itself, rather than obtaining a bootstrap estimate for each trial. The estimates obtained through either method are approximately equal, and making use of the former dramatically reduces the runtime of our algorithm. This gives us a total time complexity of $O((n_b + n_p) \cdot (n + k))$ instead of a higher time complexity of $O(n_b \cdot (1 + n_p) \cdot (n + k))$ for not much gain ($n_b$ is the number of bootstrap trials, $n_p$ is the number of permutation trials, $n$ is the sample size considered, and the statistic computation is assumed to have a time complexity of $O(k)$).
\smallskip

\noindent \textbf{Experimental Setup: }
We performed our experiments on the \textit{ProPublica} COMPAS dataset \cite{compas} (used for recidivism prediction and informing bail decisions) and the Adult dataset from the UCI Machine Learning Repository \cite{kohavi1996scaling} (used for predicting income). The COMPAS dataset contains $6167$ records, with the labels indicating whether a criminal defendant committed a crime within two years or not. The Adult dataset contains $48842$ records, with the labels specifying if an individual makes over $\$50000$ a year.

Both datasets were divided into an approximate $70\%-15\%-15\%$ train-validation-test split. We made use of all the features available except gender and race, which we treated as protected attributes. The numerical features were used as-is, while the categorical features were one-hot encoded. We also ignored the `final weight' feature in the Adult dataset. We then trained a logistic regression model with $L2$ regularization on each of these datasets, producing final models with a test AUC of $0.7005$ for the COMPAS dataset, and a test AUC of $0.9087$ for the Adult dataset.
\smallskip

\noindent \textbf{Definitions of Fairness: }
Let the classifier be defined by the function $H: X \rightarrow \{0, 1\}$, where $X$ is the input data point and the output is the predicted label. The labels of the data points $X$ are given by the function $Y: X \rightarrow \{0, 1\}$, and the protected attributes are defined by the function $G: X \rightarrow \mathcal{G}$ ($\mathcal{G}$ is the set of protected attribute values). Let $TP$ be the number of True Positives produced by the classifier and $P$ be the number of positive labels (we treat $1$ as positives and $0$ as the negative labels here). Using these notational conventions, some common metrics to assess fairness are defined below:
\begin{enumerate}
\item A classifier is said to have achieved Equalized Odds if $ \forall y \in \{0, 1\},\; g \in \mathcal{G}$,
\begin{align*}
\EE\left[H(X) = 1 | Y(X) = y, G(X) = g\right] = \EE\left[H(X) = 1 | Y(X) = y\right]
\end{align*}

Defining Equalized Odds Distances as:
\begin{align*}
\delta_{(g_1,g_2,y)} &= \EE\left[H(X) = 1 | Y(X) = y, G(X) = g_1\right] \\
&\;\;\; - \EE\left[H(X) = 1 | Y(X) = y, G(X) = g_2\right]
\end{align*}

we see that Equalized Odds can equivalently be defined as $$\delta_{(g_1,g_2,y)} = 0 \;\forall y \in \{0, 1\},\; g_1, g_2 \in \mathcal{G}$$
We thus make use of the $\delta$s as metrics for fairness.
\item Recall (or True Positive Rate, TPR) is defined as
$$\frac{TP}{P} = \EE\left[H(X) = 1 | Y(X) = 1\right]$$
Thus the difference in recall values between two protected groups $g_1$ and $g_2$ is nothing but $\delta_{(g_1, g_2, 1)}$ from above. We make use of this for performing permutation tests.
\item False Positive Rate (FPR) is defined as
$$\frac{FP}{N} = \EE\left[H(X) = 1 | Y(X) = 0\right]$$
Thus the difference in FPR values between two protected groups $g_1$ and $g_2$ is nothing but $\delta_{(g_1, g_2, 0)}$ from above. We make use of this for performing permutation tests.
\end{enumerate}

\subsection{Empirical Analyses}
The first empirical analysis compares the output of the permutation test 
with conventional fairness metrics. Specifically, we focus on performing a permutation test for the Recall (TPR) and the False Positive Rate (FPR) and compare this with the Equalized Odds fairness metric.

\begin{figure}[ht]
  \centering
  \includegraphics[width=\linewidth]{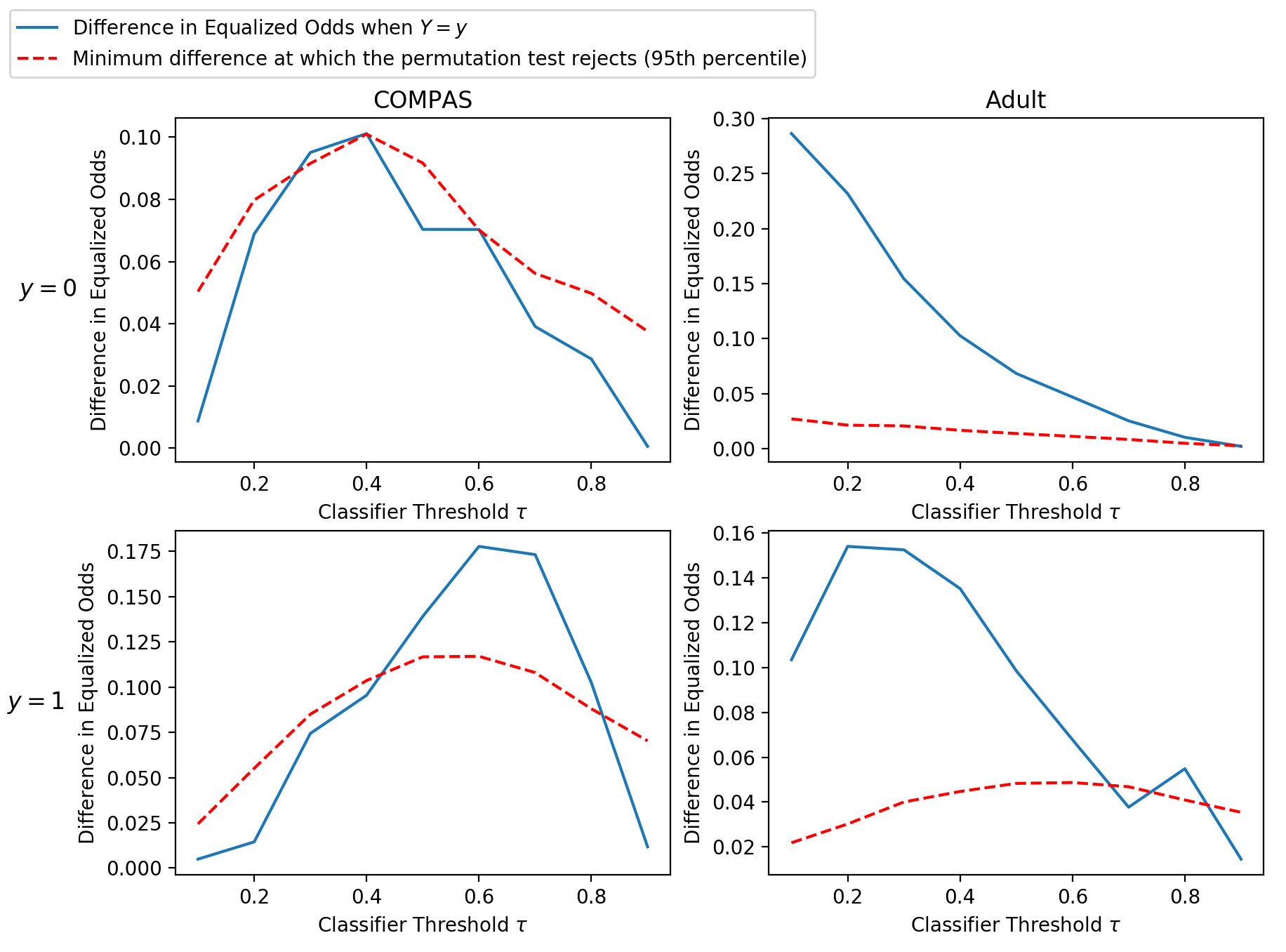}
  \caption{Comparing Equalized Odds with Permutation Testing in the context of FPR (Equalized Odds when $Y=0$) and Recall (Equalized Odds when $Y=1$), as the classifier threshold $\tau$ is varied.}
  \label{fig:reject}
\end{figure}

We consider $\mathcal{G}$ to be the gender of the individual, comprised of two elements, Male (M) and Female (F). The classifier threshold $\tau$ is varied from $0.0$ to $1.0$, and both the COMPAS (about $1000$ uniformly random samples) and Adult (about $7400$ uniformly random samples) test datasets are classified to measure $\delta_{(M, F, 0)}$ and $\delta_{(M, F, 1)}$. We also ran permutation tests for FPR and Recall for each value of $\tau$, using $1000$ permutation trials and a significance level of $0.05$ to reject the null hypothesis. Figure \ref{fig:reject} shows the resulting graphs, depicting both the Equalized Odds distance as well as the 95th percentile of the permutation distribution (values greater than this are rejected by the test for our chosen significance level).

Increasing $\tau$ reduces both the overall Recall and FPR values for the resulting classifier. When $\tau$ equals $0.0$ or $1.0$, perfect Equalized Odds is achieved due to all examples being classified as positives or negatives respectively (Recall and FPR rates are equal for all protected groups). However, intermediate values of $\tau$ result in varying degrees of Equalized Odds unfairness. It is up to the end-user to identify whether this difference is large enough to warrant addressing and whether this difference is just a statistical anomaly. However, the permutation test makes a statistically sound decision regarding this, deeming the differences to be unfair only when it crosses the 95th percentile for a given value of $\tau$.

Recall that the test statistic being computed is the difference in an aggregate metric computed for each subset of the data (resulting from a partitioning of the data into two subsets). Let $X$ and $Y$ be the resulting partitions comprised of the random variables $X_i$ and $Y_j$ respectively, each occuring in sample sizes of $n_X$ and $n_Y$. Let the aggregate metric be given by $f$, with the metric at the individual level being $f_i$. For simplicity, let us consider $f$ to be a proportion statistic, but we can extend this reasoning to other metrics as well.  It is well know that the variance of  the test statistic $T$ is $\mathcal{O}(1/\min \{n_x, n_Y\})$.
Hence, as the sample sizes $n_X$ and $n_Y$ increase the variance of $T$ decreases. Consequently, as the sample size increases, the permutation test is able to reject the null hypothesis for smaller differences at a fixed significance level.

\begin{figure}[ht]
\begin{center}
\centerline{\includegraphics[width=0.7\linewidth]{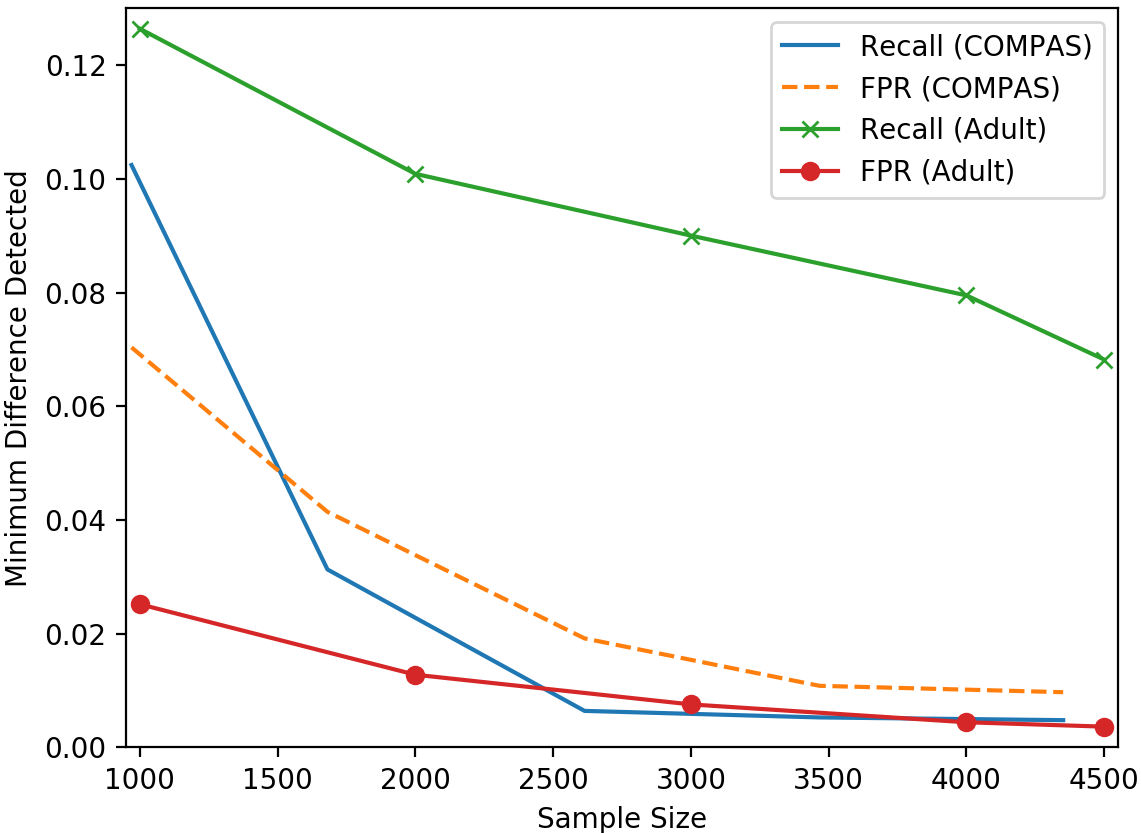}}
\caption{Minimum Difference Detected vs Sample Size}
\label{fig:detected_diff}
\end{center}
\vspace{-0.5cm}
\end{figure}

In the second experiment, we looked into the effect of sample size on permutation testing, varying it from around $950$ to $4500$. For the Adult dataset, we were able to sample the test dataset directly, but for the COMPAS dataset, we had to take uniformly random samples from the training data due to insufficient test data points. For each choice of sample size, we varied $\tau$ from $0.0$ to $1.0$ to obtain the minimum difference in Recall and FPR that was detected by the permutation test at a significance level of $0.05$, with each test being run for $1000$ permutation trials. Figure \ref{fig:detected_diff} shows the resulting graph, from which we can conclude that an increase in the sample size allows smaller differences to be detected with statistical significance.

Conversely, given a minimum difference to be detected, one can compute the sample size to be used for the permutation test. For example, suppose that $n_X = n_Y$, the test statistic $T$ follows a normal distribution, and we wish to identify the sample size $n_X + n_Y$ such that the permutation test rejects at a significance level of $0.05$. Since $2 \sqrt{\text{var}\left(T\right)}$ is approximately the $95$th percentile, if we can estimate $\text{var}\left(f_i\left(X_i\right)\right)$, we can use the equation for $\text{var}\left(T\right)$ to obtain an estimate for $n_X$ and $n_Y$. As an example, working with the false negative rate example from Section~\ref{sec:proportion-stats}, we can sample multiple $X_i$, score them with the model and check if it is a false negative, thereby providing us with samples of the indicator function to estimate $\text{var}\left(f_i\left(X_i\right)\right)$ with.

Another effect of increasing the sample size is a reduction in the standard error of the p-value estimation (and consequently, a smaller confidence interval). Under the null hypothesis, the permutation test can be treated as $n$ statistically independent trials for which the probability of observing an extreme result remains the same. Thus, we can estimate the p-value as a binomial proportion, dividing the number of extreme trials (those resulting in values at least as extreme as the observed difference) by the total number of trials. We can also estimate the confidence interval and standard deviation of our estimate by making use of techniques such as those described in \citep{agresti1998approximate, wilson1927probable}. There is also work \citep{brown2001interval} that compares these estimates and makes recommendations, but the common factor between these estimates is that they are inversely proportional to the number of trials raised to some power, indicating that our estimates of the p-value have lower standard errors and smaller confidence intervals if the number of trials is increased.

\section{Related Work}
\label{sec:related}
 There is extensive literature on different notions of fairness. \cite{speicher2018unified} proposes using inequality indices from economics, namely the \textit{Generalized Entropy Index} (GEI), to measure how model predictions unequally benefit different groups. It also allows for the decomposition of the fairness metric into between-group and within-group components, to better understand where the source of inequality truly lies. \textit{Conditional Equality of Opportunity} is another metric, proposed in \cite{beutel2019putting} as a technique to account for distributional differences. This builds on the notion of Conditional Parity \cite{ritov2017conditional}, which discusses fairness constraints conditioned on certain attributes as more general notions of fairness criteria like Demographic Parity and Equalized Odds. Conditional Equality quantifies this fairness criterion as a weighted sum (over the conditional attribute values) of individual conditional attribute deviations, with the weights being defined by how much importance certain attribute values are to be given over others. Although these metrics quantify the amount of `unfairness' in an algorithm, they deem any non-zero value to be `unfair'. These metrics by themselves are insufficient to declare an algorithm to be unfair; we need statistically sound techniques, such as permutation tests, to reject the null hypothesis of fairness.  
There are numerous open-source packages computing fairness metrics including IBM's AI Fairness 360\footnote{https://aif360.mybluemix.net}, Google's ML-Fairness-Gym \cite{damour20}, Themis \cite{galhotra2017}, and FairTest  \cite{tramer2017}, though many do not incorporate formal hypothesis testing. 
 Permutation tests have been used to assess the performance of predictive models (e.g. \cite{ojala2009}). Further, robust permutation tests for two-sample problems have been proposed in \cite{chung2013}. We are not aware of any related work that established the validity of permutation testing for assessing fairness.
\vspace{-0.1cm}
\section{Conclusion}
\label{sec:conclusion}
There are many aspects of algorithmic fairness that are captured by various metrics and definitions.  No single metric captures all aspects of fairness, and we would encourage a practitioner to evaluate fairness along multiple metrics to better understand where biases may be present.  For this purpose, our contribution is to provide a methodology to assess the strength of evidence that a model may be unfair with respect to any metric a researcher may be interested in.  The framework for permutation testing proposed in this paper provides a flexible, non-parametric approach to assessing fairness, thereby simplifying the burden of performing a statistical test on the practitioner to merely specifying a test statistic. Moreover, the framework attempts to close the gap of not having a formal statistical test for detecting unfairness. 

We demonstrated the performance of permutation testing through extensive experiments on two real-world datasets known to exhibit bias.  
An interesting aspect of the simulation result is that a classifier exhibits bias for differing values of a threshold.   Moreover, the values of the threshold for which bias was detectable depended on the metric under consideration.  This reinforces the need to experiment with multiple definitions of fairness while attempting to determine if a model is biased.  Testing across multiple metrics is greatly simplified by the use of our non-parametric testing framework. 
We also showed that our framework provides a better distributional approximation than the bootstrap.

Although the discussion in this paper focused on binary classification problems, mainly for simplicity of exposition and brevity, we remark that our methodology is also applicable to most other supervised learning problem settings.


\bibliographystyle{ACM-Reference-Format}  
\bibliography{paper}


\begin{thebibliography}{26}


\ifx \showCODEN    \undefined \def \showCODEN     #1{\unskip}     \fi
\ifx \showDOI      \undefined \def \showDOI       #1{#1}\fi
\ifx \showISBNx    \undefined \def \showISBNx     #1{\unskip}     \fi
\ifx \showISBNxiii \undefined \def \showISBNxiii  #1{\unskip}     \fi
\ifx \showISSN     \undefined \def \showISSN      #1{\unskip}     \fi
\ifx \showLCCN     \undefined \def \showLCCN      #1{\unskip}     \fi
\ifx \shownote     \undefined \def \shownote      #1{#1}          \fi
\ifx \showarticletitle \undefined \def \showarticletitle #1{#1}   \fi
\ifx \showURL      \undefined \def \showURL       {\relax}        \fi
\providecommand\bibfield[2]{#2}
\providecommand\bibinfo[2]{#2}
\providecommand\natexlab[1]{#1}
\providecommand\showeprint[2][]{arXiv:#2}

\bibitem[\protect\citeauthoryear{Agarwal, Beygelzimer, Dudik, Langford, and
  Wallach}{Agarwal et~al\mbox{.}}{2018}]%
        {agarwal2018reductions}
\bibfield{author}{\bibinfo{person}{Alekh Agarwal}, \bibinfo{person}{Alina
  Beygelzimer}, \bibinfo{person}{Miroslav Dudik}, \bibinfo{person}{John
  Langford}, {and} \bibinfo{person}{Hanna Wallach}.}
  \bibinfo{year}{2018}\natexlab{}.
\newblock \showarticletitle{A Reductions Approach to Fair Classification}. In
  \bibinfo{booktitle}{\emph{International Conference on Machine Learning}}.
  \bibinfo{pages}{60--69}.
\newblock


\bibitem[\protect\citeauthoryear{Agresti and Coull}{Agresti and Coull}{1998}]%
        {agresti1998approximate}
\bibfield{author}{\bibinfo{person}{Alan Agresti} {and} \bibinfo{person}{Brent~A
  Coull}.} \bibinfo{year}{1998}\natexlab{}.
\newblock \showarticletitle{Approximate is better than "exact" for interval
  estimation of binomial proportions}.
\newblock \bibinfo{journal}{\emph{The American Statistician}}
  \bibinfo{volume}{52}, \bibinfo{number}{2} (\bibinfo{year}{1998}),
  \bibinfo{pages}{119--126}.
\newblock


\bibitem[\protect\citeauthoryear{Angwin, Larson, Mattu, and Kirchner}{Angwin
  et~al\mbox{.}}{2016}]%
        {angwin_2016}
\bibfield{author}{\bibinfo{person}{Julia Angwin}, \bibinfo{person}{Jeff
  Larson}, \bibinfo{person}{Surya Mattu}, {and} \bibinfo{person}{Lauren
  Kirchner}.} \bibinfo{year}{2016}\natexlab{}.
\newblock \showarticletitle{Machine bias}.
\newblock \bibinfo{journal}{\emph{ProPublica}} (\bibinfo{year}{2016}).
\newblock


\bibitem[\protect\citeauthoryear{{Apache Spark Team}}{{Apache Spark
  Team}}{2014}]%
        {apacheSpark}
\bibfield{author}{\bibinfo{person}{{Apache Spark Team}}.}
  \bibinfo{year}{2014}\natexlab{}.
\newblock \bibinfo{title}{Apache {S}park: {A} fast and general engine for
  large-scale data processing}.
\newblock
\newblock
\newblock
\shownote{\url{https://spark.apache.org}, Last accessed on 2019-09-10.}


\bibitem[\protect\citeauthoryear{Barocas and Hardt}{Barocas and Hardt}{2017}]%
        {nips_2017_tutorial}
\bibfield{author}{\bibinfo{person}{Solon Barocas} {and} \bibinfo{person}{Moritz
  Hardt}.} \bibinfo{year}{2017}\natexlab{}.
\newblock \showarticletitle{Fairness in Machine Learning}. In
  \bibinfo{booktitle}{\emph{NIPS Tutorial}}.
\newblock


\bibitem[\protect\citeauthoryear{Beutel, Chen, Doshi, Qian, Woodruff, Luu,
  Kreitmann, Bischof, and Chi}{Beutel et~al\mbox{.}}{2019}]%
        {beutel2019putting}
\bibfield{author}{\bibinfo{person}{Alex Beutel}, \bibinfo{person}{Jilin Chen},
  \bibinfo{person}{Tulsee Doshi}, \bibinfo{person}{Hai Qian},
  \bibinfo{person}{Allison Woodruff}, \bibinfo{person}{Christine Luu},
  \bibinfo{person}{Pierre Kreitmann}, \bibinfo{person}{Jonathan Bischof}, {and}
  \bibinfo{person}{Ed~H Chi}.} \bibinfo{year}{2019}\natexlab{}.
\newblock \showarticletitle{Putting fairness principles into practice:
  Challenges, metrics, and improvements}.
\newblock \bibinfo{journal}{\emph{arXiv preprint arXiv:1901.04562}}
  (\bibinfo{year}{2019}).
\newblock


\bibitem[\protect\citeauthoryear{Bolukbasi, Chang, Zou, Saligrama, and
  Kalai}{Bolukbasi et~al\mbox{.}}{2016}]%
        {bolukbasi_2016}
\bibfield{author}{\bibinfo{person}{Tolga Bolukbasi}, \bibinfo{person}{Kai-Wei
  Chang}, \bibinfo{person}{James~Y Zou}, \bibinfo{person}{Venkatesh Saligrama},
  {and} \bibinfo{person}{Adam~T Kalai}.} \bibinfo{year}{2016}\natexlab{}.
\newblock \showarticletitle{{Man is to computer programmer as woman is to
  homemaker? Debiasing word embeddings}}. In \bibinfo{booktitle}{\emph{NIPS}}.
\newblock


\bibitem[\protect\citeauthoryear{Brown, Cai, and DasGupta}{Brown
  et~al\mbox{.}}{2001}]%
        {brown2001interval}
\bibfield{author}{\bibinfo{person}{Lawrence~D Brown}, \bibinfo{person}{T~Tony
  Cai}, {and} \bibinfo{person}{Anirban DasGupta}.}
  \bibinfo{year}{2001}\natexlab{}.
\newblock \showarticletitle{Interval estimation for a binomial proportion}.
\newblock \bibinfo{journal}{\emph{Statistical science}} (\bibinfo{year}{2001}),
  \bibinfo{pages}{101--117}.
\newblock


\bibitem[\protect\citeauthoryear{Caliskan, Bryson, and Narayanan}{Caliskan
  et~al\mbox{.}}{2017}]%
        {caliskan_2017}
\bibfield{author}{\bibinfo{person}{Aylin Caliskan}, \bibinfo{person}{Joanna~J
  Bryson}, {and} \bibinfo{person}{Arvind Narayanan}.}
  \bibinfo{year}{2017}\natexlab{}.
\newblock \showarticletitle{Semantics derived automatically from language
  corpora contain human-like biases}.
\newblock \bibinfo{journal}{\emph{Science}} \bibinfo{volume}{356},
  \bibinfo{number}{6334} (\bibinfo{year}{2017}).
\newblock


\bibitem[\protect\citeauthoryear{Chung and Romano}{Chung and Romano}{2013}]%
        {chung2013}
\bibfield{author}{\bibinfo{person}{EunYi Chung} {and}
  \bibinfo{person}{Joseph~P. Romano}.} \bibinfo{year}{2013}\natexlab{}.
\newblock \showarticletitle{Exact and asymptotically robust permutation tests}.
\newblock \bibinfo{journal}{\emph{Ann. Statist.}} \bibinfo{volume}{41},
  \bibinfo{number}{2} (\bibinfo{date}{04} \bibinfo{year}{2013}),
  \bibinfo{pages}{484--507}.
\newblock


\bibitem[\protect\citeauthoryear{Chung and Romano}{Chung and Romano}{2016}]%
        {chung2016}
\bibfield{author}{\bibinfo{person}{EunYi Chung} {and}
  \bibinfo{person}{Joseph~P. Romano}.} \bibinfo{year}{2016}\natexlab{}.
\newblock \showarticletitle{Asymptotically valid and exact permutation tests
  based on two-sample U-statistics}.
\newblock \bibinfo{journal}{\emph{Journal of Statistical Planning and
  Inference}}  \bibinfo{volume}{168} (\bibinfo{year}{2016}), \bibinfo{pages}{97
  -- 105}.
\newblock


\bibitem[\protect\citeauthoryear{Dwork, Hardt, Pitassi, and
  Omer~Reingold}{Dwork et~al\mbox{.}}{2012}]%
        {dwork_2012}
\bibfield{author}{\bibinfo{person}{Cynthia Dwork}, \bibinfo{person}{Moritz
  Hardt}, \bibinfo{person}{Toniann Pitassi}, {and}
  \bibinfo{person}{Richard~Zemel Omer~Reingold}.}
  \bibinfo{year}{2012}\natexlab{}.
\newblock \showarticletitle{Fairness through awareness}. In
  \bibinfo{booktitle}{\emph{ITCS}}.
\newblock


\bibitem[\protect\citeauthoryear{D’Amour, Srinivasan, Atwood, Baljekar,
  Sculley, and Halpern}{D’Amour et~al\mbox{.}}{2020}]%
        {damour20}
\bibfield{author}{\bibinfo{person}{Alexander D’Amour}, \bibinfo{person}{Hansa
  Srinivasan}, \bibinfo{person}{James Atwood}, \bibinfo{person}{Pallavi
  Baljekar}, \bibinfo{person}{D. Sculley}, {and} \bibinfo{person}{Yoni
  Halpern}.} \bibinfo{year}{2020}\natexlab{}.
\newblock \showarticletitle{Fairness is Not Static: Deeper Understanding of
  Long Term Fairness via Simulation Studies}. In
  \bibinfo{booktitle}{\emph{Proceedings of the 2020 Conference on Fairness,
  Accountability, and Transparency}} \emph{(\bibinfo{series}{FAT* ’20})}.
  \bibinfo{publisher}{Association for Computing Machinery},
  \bibinfo{pages}{525–534}.
\newblock


\bibitem[\protect\citeauthoryear{Galhotra, Brun, and Meliou}{Galhotra
  et~al\mbox{.}}{2017}]%
        {galhotra2017}
\bibfield{author}{\bibinfo{person}{Sainyam Galhotra}, \bibinfo{person}{Yuriy
  Brun}, {and} \bibinfo{person}{Alexandra Meliou}.}
  \bibinfo{year}{2017}\natexlab{}.
\newblock \showarticletitle{Fairness Testing: Testing Software for
  Discrimination}. In \bibinfo{booktitle}{\emph{Proceedings of the 2017 11th
  Joint Meeting on Foundations of Software Engineering}}.
  \bibinfo{publisher}{Association for Computing Machinery},
  \bibinfo{pages}{498–510}.
\newblock


\bibitem[\protect\citeauthoryear{Good}{Good}{2000}]%
        {good2000permutation}
\bibfield{author}{\bibinfo{person}{P.I. Good}.}
  \bibinfo{year}{2000}\natexlab{}.
\newblock \bibinfo{booktitle}{\emph{Permutation Tests: A Practical Guide to
  Resampling Methods for Testing Hypotheses}}.
\newblock \bibinfo{publisher}{Springer}.
\newblock


\bibitem[\protect\citeauthoryear{Hajian, Bonchi, and Castillo}{Hajian
  et~al\mbox{.}}{2016}]%
        {hajian_2016_tutorial}
\bibfield{author}{\bibinfo{person}{S. Hajian}, \bibinfo{person}{F. Bonchi},
  {and} \bibinfo{person}{C. Castillo}.} \bibinfo{year}{2016}\natexlab{}.
\newblock \showarticletitle{Algorithmic Bias: From Discrimination Discovery to
  Fairness-aware Data Mining}. In \bibinfo{booktitle}{\emph{KDD Tutorial on
  Algorithmic Bias}}.
\newblock


\bibitem[\protect\citeauthoryear{Kay, Matuszek, and Munson}{Kay
  et~al\mbox{.}}{2015}]%
        {kay_2015}
\bibfield{author}{\bibinfo{person}{Matthew Kay}, \bibinfo{person}{Cynthia
  Matuszek}, {and} \bibinfo{person}{Sean~A. Munson}.}
  \bibinfo{year}{2015}\natexlab{}.
\newblock \showarticletitle{Unequal Representation and Gender Stereotypes in
  Image Search Results for Occupations}. In \bibinfo{booktitle}{\emph{CHI}}.
\newblock


\bibitem[\protect\citeauthoryear{Kohavi}{Kohavi}{1996}]%
        {kohavi1996scaling}
\bibfield{author}{\bibinfo{person}{Ron Kohavi}.}
  \bibinfo{year}{1996}\natexlab{}.
\newblock \showarticletitle{Scaling up the accuracy of {Naive-Bayes}
  classifiers: A decision-tree hybrid}. In \bibinfo{booktitle}{\emph{KDD}}.
\newblock


\bibitem[\protect\citeauthoryear{Larson, Mattu, Kirchner, and Angwin}{Larson
  et~al\mbox{.}}{2016}]%
        {compas}
\bibfield{author}{\bibinfo{person}{Jeff Larson}, \bibinfo{person}{Surya Mattu},
  \bibinfo{person}{Lauren Kirchner}, {and} \bibinfo{person}{Julia Angwin}.}
  \bibinfo{year}{2016}\natexlab{}.
\newblock \bibinfo{title}{Data and analysis for `How we analyzed the COMPAS
  recidivism algorithm'}.
\newblock
\newblock
\newblock
\shownote{\url{https://github.com/propublica/compas-analysis}, Last accessed on
  2019-09-10.}


\bibitem[\protect\citeauthoryear{Mary, Calauzenes, and El~Karoui}{Mary
  et~al\mbox{.}}{2019}]%
        {mary2019fairness}
\bibfield{author}{\bibinfo{person}{J{\'e}r{\'e}mie Mary},
  \bibinfo{person}{Cl{\'e}ment Calauzenes}, {and} \bibinfo{person}{Noureddine
  El~Karoui}.} \bibinfo{year}{2019}\natexlab{}.
\newblock \showarticletitle{Fairness-aware learning for continuous attributes
  and treatments}. In \bibinfo{booktitle}{\emph{International Conference on
  Machine Learning}}. \bibinfo{pages}{4382--4391}.
\newblock


\bibitem[\protect\citeauthoryear{Ojala and Garriga}{Ojala and Garriga}{2010}]%
        {ojala2009}
\bibfield{author}{\bibinfo{person}{Markus Ojala} {and}
  \bibinfo{person}{Gemma~C. Garriga}.} \bibinfo{year}{2010}\natexlab{}.
\newblock \showarticletitle{Permutation Tests for Studying Classifier
  Performance}.
\newblock \bibinfo{journal}{\emph{Journal of Machine Learning Research}}
  \bibinfo{volume}{11} (\bibinfo{year}{2010}), \bibinfo{pages}{1833--1863}.
\newblock


\bibitem[\protect\citeauthoryear{Pedreschi, Ruggieri, and Turini}{Pedreschi
  et~al\mbox{.}}{2009}]%
        {pedreschi_2009}
\bibfield{author}{\bibinfo{person}{Dino Pedreschi}, \bibinfo{person}{Salvatore
  Ruggieri}, {and} \bibinfo{person}{Franco Turini}.}
  \bibinfo{year}{2009}\natexlab{}.
\newblock \showarticletitle{Measuring discrimination in socially-sensitive
  decision records}. In \bibinfo{booktitle}{\emph{SDM}}.
\newblock


\bibitem[\protect\citeauthoryear{Ritov, Sun, and Zhao}{Ritov
  et~al\mbox{.}}{2017}]%
        {ritov2017conditional}
\bibfield{author}{\bibinfo{person}{Ya'acov Ritov}, \bibinfo{person}{Yuekai
  Sun}, {and} \bibinfo{person}{Ruofei Zhao}.} \bibinfo{year}{2017}\natexlab{}.
\newblock \showarticletitle{On conditional parity as a notion of
  non-discrimination in machine learning}.
\newblock \bibinfo{journal}{\emph{arXiv preprint arXiv:1706.08519}}
  (\bibinfo{year}{2017}).
\newblock


\bibitem[\protect\citeauthoryear{Speicher, Heidari, Grgic-Hlaca, Gummadi,
  Singla, Weller, and Zafar}{Speicher et~al\mbox{.}}{2018}]%
        {speicher2018unified}
\bibfield{author}{\bibinfo{person}{Till Speicher}, \bibinfo{person}{Hoda
  Heidari}, \bibinfo{person}{Nina Grgic-Hlaca}, \bibinfo{person}{Krishna~P
  Gummadi}, \bibinfo{person}{Adish Singla}, \bibinfo{person}{Adrian Weller},
  {and} \bibinfo{person}{Muhammad~Bilal Zafar}.}
  \bibinfo{year}{2018}\natexlab{}.
\newblock \showarticletitle{A Unified Approach to Quantifying Algorithmic
  Unfairness: Measuring Individual \&Group Unfairness via Inequality Indices}.
  In \bibinfo{booktitle}{\emph{Proceedings of the 24th ACM SIGKDD International
  Conference on Knowledge Discovery \& Data Mining}}. ACM,
  \bibinfo{pages}{2239--2248}.
\newblock


\bibitem[\protect\citeauthoryear{{Tramèr}, {Atlidakis}, {Geambasu}, {Hsu},
  {Hubaux}, {Humbert}, {Juels}, and {Lin}}{{Tramèr} et~al\mbox{.}}{2017}]%
        {tramer2017}
\bibfield{author}{\bibinfo{person}{F. {Tramèr}}, \bibinfo{person}{V.
  {Atlidakis}}, \bibinfo{person}{R. {Geambasu}}, \bibinfo{person}{D. {Hsu}},
  \bibinfo{person}{J. {Hubaux}}, \bibinfo{person}{M. {Humbert}},
  \bibinfo{person}{A. {Juels}}, {and} \bibinfo{person}{H. {Lin}}.}
  \bibinfo{year}{2017}\natexlab{}.
\newblock \showarticletitle{FairTest: Discovering Unwarranted Associations in
  Data-Driven Applications}. In \bibinfo{booktitle}{\emph{2017 IEEE European
  Symposium on Security and Privacy (EuroS P)}}. \bibinfo{pages}{401--416}.
\newblock
\showISSN{null}


\bibitem[\protect\citeauthoryear{Wilson}{Wilson}{1927}]%
        {wilson1927probable}
\bibfield{author}{\bibinfo{person}{Edwin~B Wilson}.}
  \bibinfo{year}{1927}\natexlab{}.
\newblock \showarticletitle{Probable inference, the law of succession, and
  statistical inference}.
\newblock \bibinfo{journal}{\emph{J. Amer. Statist. Assoc.}}
  \bibinfo{volume}{22}, \bibinfo{number}{158} (\bibinfo{year}{1927}),
  \bibinfo{pages}{209--212}.
\newblock


\end{thebibliography}

\section*{Appendix}
\label{sec:supplement}
Here we present the proofs of the results in the main text. 

\subsection{Proof of Theorem \ref{thm:AUC}}
To find the asymptotic distribution of the difference of AUC's, we will first find the asymptotic distribution of 
\begin{align*}
\widehat{AUC}_A  & = \frac{\sqrt{n}}{n^+_An^-_A} \sum_{i,j} I \left\{ f(X_{A,i}^+) > f(X_{A,j}^-) \right\} \\
&= \frac{\sqrt{n}}{n^+_An^-_A}  \sum_{i,j} I \left\{ f(X_{A,i}) > f(X_{A,j}), y_{i,A} = +1, y_{j,A} = -1 \right\}.
\end{align*}
Write $AUC$ for the common AUC of group A and B.  Define $$\kappa_A(d_{i,A},d_{j,A}) = \frac{t(d_{i,A} ,d_{j,A}) + t(d_{j,A} ,d_{i,A})}{2}$$ with $d_{i,A} = (X_{i,A}, y_{i,A})$ and 
$$ t(d_i, d_j) = I \left\{ f(X_{A,i}) > f(X_{A,j}), y_{i,A} = +1, y_{j,A} = -1 \right\}. $$
Then, the multivariate central limit theorem for U statistics gives 
\begin{align*}
\sqrt{n_A} &\left( \begin{matrix} 
\frac{2}{n_A (n_A-1)} \sum_{i,j} \kappa_A(d_{i,A}, d_{j,A}) - AUC p_{+,A}(1-p_{+,A}) \\
\frac{1}{n_A}\sum_{i} I \left\{ y_{i,A} = +1\right\} - p_{+,A}
\end{matrix} \right) \\
&\qquad \qquad \qquad \qquad \qquad \qquad \qquad \qquad \qquad \rightarrow N(0, \Sigma_A)
\end{align*}
where $\Sigma_{A}$ has entries 
$$(\Sigma_{A})_{1,1} = p_{+,A} (1-p_{+,A}),$$

$$(\Sigma_{A})_{2,1} = AUC p_{+,A} (1 - p_{+,A})(1 - 2p_{+,A}),$$
and
\begin{align*}
(\Sigma_{A})_{1,1} = 
	p_{+,A} (1 - p_{+,A}) &[ p_{+--} (1 - p_{+,A}) + p_{++-} p_{+,A} \\
	& \;\; - 4AUC p_{+,A} (1 - p_{+,A})] 
\end{align*}
with 
$$p_{+--} = P(f(X_{A,1}^+) > f(X_{A,1}^-), f(X_{A,1}^+) > f(X_{A,2}^-))$$
and 
$$p_{++-} = P(f(X_{A,1}^+) > f(X_{A,1}^-),f(X_{A,2}^+) > f(X_{A,1}^-)).$$ 
Applying the delta method to $f(x,y) = x/(y(1-y))$, which has relevant gradient 
\begin{align*}
&\nabla f (AUC p_{+,A}(1-p_{+,A}), p_{+,A}) \\
&\qquad \qquad = \frac{1}{p_{+,A}(1-p_{+,A})} \left(1, AUC (2 p_{+,A} -1) \right)
\end{align*}
gives that 
$$\sqrt{n}\left( \widehat{AUC}_A - AUC \right) \rightarrow N(0,v_A/p_A).$$
Noting that the estimated AUC for group B is independent of that of group A yields the limiting distribution for the sampling distribution given in Theorem \ref{thm:AUC}.

To derive the asymptotic distribution of the permutation distribution, write $d_1,...,d_n = d_{1,A},...,d_{n_A,A},d_{1,B},...,d_{n_B,B}$.  The obvious multivariate extension of the limiting results for the permutation distribution of U-statistics given in \cite{chung2016} yields that the permutation distribution of 
 $$\sqrt{n} \left( \begin{matrix} 
	\frac{2}{n_A (n_A-1)} \sum_{\pi(i),\pi(j) \leq n_A} \phi(d_{\pi(i)}, d_{\pi(j)}) - AUC p_{+}(1-p_{+})\\
	\frac{1}{n_A}\sum_{\pi(i) \leq n_A} I \left\{ y_{i} = +1\right\} - p_{+}\\
	\frac{2}{n_B (n_B-1)} \sum_{\pi(i),\pi(j) > n_A} \phi(d_{pi(i)}, d_{pi(j)}) - AUC p_{+}(1-p_{+})\\
	\frac{1}{n_B}\sum_{\pi(i) > n_A} I \left\{ y_{\pi(i)} = +1\right\} - p_{+}
\end{matrix} \right)$$
is asymptotically normal, in probability, with mean $0$ and variance 
$$\Sigma_\pi = \left( \begin{matrix} 
\frac{1}{p_A} \Sigma &0 \\
0 & \frac{1}{p_B} \Sigma
\end{matrix} \right)$$
in probability, where $\Sigma$ is given by the same expression as $\Sigma_A$ had group A been sampled from distribution $p_A P_{(X_A,y_A)} +(1-p_A) P_{(X_B,y_B) }$.  Following the same delta method calculation as for the sampling distribution gives the desired asymptotic distribution of the permutation distribution.

\subsection{Proof of Theorem \ref{thm:AUCstudentized}}
The results of Theorem \ref{thm:AUCstudentized} follow immediately from Slutsky's Theorem.

\subsection{Derivation of limiting distribution of FNR}

For further compactness, write $c_{i,A} = c(X_{i,A})$.  The false negative rate of group A can be written as 
$$\hat p^-_A = \frac{\sum_{i+1}^{n_A} I \left\{ c_{i,A} = -1, y_{i,A} = 1\right\}}{\sum_{i+1}^{n_A} I \left\{ y_{i,A} = 1\right\}} = \frac{N_A}{D_A}$$
where we define $N_A$ and $D_A$ to  be the numerator and denominator of the proceeding quantity.  Define $N_B$ and $D_B$ analogously for group B.  We wish to study the limiting behavior of 
$$\sqrt{n} \left( \frac{N_A}{D_A} - \frac{N_B}{D_B} \right) = \sqrt{n} \left( \frac{N_A}{D_A} - p_{FP} \right) - \sqrt{n} \left( \frac{N_A}{D_A} - p_{FP} \right).$$
Since group A and group B are independent, it is enough to establish  the asymptotic normality of  
$$\sqrt{n_A} \left( \frac{N_A}{D_A} - p_{FN} \right)$$
(and the same quantity for group B).  Finding the limiting distribution is a routine application of the delta method.  Assume $n_A/(n_A+n_B) = p_A \in (0,1)$ and $EI\left\{y_{i,A} = +1\right\} = p_{+, A} \in(0,1)$.
Then, $$\sqrt{n_A} \left( \begin{matrix} 
N_A - p_{+,A} p_{FN} \\
N_B - p_{+,A}
\end{matrix} \right) \rightarrow N(0, \Sigma_A)$$
where 
$$\Sigma_A = \left( \begin{matrix} 
p_{+,A} p_{FN} (1 - p_{+,A} p_{FN}) & p_{+,A} p_{FN}(1-p_{+,A}) \\
p_{+,A} p_{FN}(1-p_{+,A}) & p_{+,A} (1-p_{+,A})
\end{matrix} \right).$$
Applying the delta method with the function $f(x,y) = x/y$ (which has gradient $\nabla f = (1/y, -x/y^2)^T$) gives that 
$$\sqrt{n_A} \left( \frac{N_A}{D_A} - p_{FN} \right) \rightarrow N(0, \Sigma_{FN_A})$$
where 
$$\Sigma_{FN_A} = \nabla f(p_{+,A} p_{FN}, p_{+,A})^T \Sigma_A \nabla f(p_{+,A} p_{FN}, p_{+,A}).$$
Simple matrix algebra yields 
$$\Sigma_{FN_A} = \frac{p_{FN}(1-p_{FN})}{p_{+,A}}.$$
Assuming $n_a/n = p_A \in (0,1)$, Slutsky's theorem gives 
$$\sqrt{n} \left( \frac{N_A}{D_A} - p_{FN} \right) \rightarrow N(0, p_{FN}(1-p_{FN})/(p_A p_{+,A})).$$

\subsection{Proof of Theorem \ref{thm:FNR}}
Suppose that $\pi$ is a uniformly chosen permutation. We wish to study the asymptotic behavior of $T_\pi$, the statistic computed on the permuted data.  

Suppose that $(X_1,y_1),...,(X_n,y_n)$ is the combined feature and label data for groups A and B indexed in no particular order and $c_1,...,c_n$ are the corresponding classifications.  We can write the difference of false negative proportions (scaled by $\sqrt{n}$) as
$$T_{\pi} =\sqrt{n} \left( \frac{\sum_{i=1}^n a_{i}b_{\pi_k(i)}}{\sum_{i=1}^n d_{i}b_{\pi_k(i)}} - \frac{\sum_{i=1}^n a_{i}b'_{\pi_k(i)}}{\sum_{i=1}^n d_{i}b_{\pi_k(i)}'} \right)$$
where 
$$a_i = I \left\{ y_i = 1, c_i = -1\right\},$$
$$d_i = I \left\{ y_i = 1\right\},$$,
$$b_i = \left\{  \begin{matrix}
1 \text{ if }i \leq n_A \\
0 \text{ otherwise}
\end{matrix} \right. ,
$$
and $b_i' = 1-b_i$.

Let $S$ be the set of observations satisfying the following conditions.
\begin{itemize}
	\item $(S1):$ $\frac{1}{n} \sum^n_{i=} I \left\{ y_i = 1 \right\}\rightarrow p_+$
	\item $(S2):$ $\frac{1}{n} \sum^n_{i=} I \left\{ y_i = 1, c_i = -1 \right\}\rightarrow p_+ p_{FN}$
\end{itemize}

We begin by deriving the distribution of $T_\pi$ conditional on $S$.  Write 
$$T_{\pi} =\sqrt{n} \left( \frac{\sum_{i=1}^n a_{i}b_{\pi(i)}}{\sum_{i=1}^n d_{i}b_{\pi(i)}} - \frac{\bar a \cdot \bar b}{\bar d \cdot \bar b} \right) - \sqrt{n} \left( \frac{\sum_{i=1}^n a_{i}b'_{\pi(i)}}{\sum_{i=1}^n d_{i}b_{\pi(i)}'}  - \frac{\bar a \cdot \bar b'}{\bar d \cdot \bar b'}\right).$$
It is readily seen using the multivariate extension of Hoeffding's combinatorial central limit theorem that 

$$\left( \begin{matrix} 
\frac{1}{\sqrt{n}} \left( \sum_{i=1}^n a_{i}b_{\pi_k(i)} - \bar a \cdot \bar b \right) \\
\frac{1}{\sqrt{n}} \left(\sum_{i=1}^n d_{i}b_{\pi_k(i)} - \bar d \cdot \bar b  \right) \\
\frac{1}{\sqrt{n}} \left(\sum_{i=1}^n a_{i}b'_{\pi_k(i)} - \bar a \cdot \bar b' \right) \\
\frac{1}{\sqrt{n}} \left( \sum_{i=1}^n d_{i}b_{\pi_k(i)}'  - \bar d \cdot \bar b' \right) \\
\end{matrix} \right) \rightarrow N(0, \Sigma_\pi)$$
conditionally on $S$, where 
$$ \Sigma_\pi = p_A(1-p_A) \left( \begin{matrix} 
V & -V \\
-V & V
\end{matrix} \right) 
$$
with 
$$ \Sigma_\pi = \left( \begin{matrix} 
p_+ p_{FN} (1-p_+ p_{FN})& p_+ p_{FN} (1- p_{FN}) \\
p_+ p_{FN} (1- p_{FN}) & p_+ (1-p_+)
\end{matrix} \right).$$
Because, $\bar a \cdot \bar b \rightarrow p_A p_+ p_{FN}$, $\bar d \cdot \bar b \rightarrow p_A p_+$, $\bar a \cdot \bar b' \rightarrow (1-p_A) p_+ p_{FN}$, and $\bar d \cdot \bar b \rightarrow (1-p_A) p_+$, at a suitable rate, the distribution of $T_{\pi_1}$ can be found using the delta method with function 
$$f(x_1,y_1, x_2, y_2) = \frac{x_1}{y_1} - \frac{x_2}{y_2}.$$
In this case, the relevant gradient is 
\begin{align*}
&\nabla f (p_A p_+ p_{FN} ,  p_A p_+,(1-p_A) p_+ p_{FN} ,  (1-p_A) p_+)^T \\
& \;\; = \left( \frac{1}{p_+ p_A} , \frac{-p_{FN}}{p_+ p_A}, \frac{-1}{p_+ (1-p_A)}, \frac{p_{FN}}{p_+ (1-p_A)}\right)
\end{align*}
An easy calculation gives 
$$\nabla f' \Sigma_{\pi} \nabla f= \frac{p_{FN}(1-p_{FN})}{p_A p_+} - \frac{p_{FN}(1-p_{FN})}{(1-p_A) p_+} .$$
It follows immediately from Slutsky's theorem applied conditionally that the studentized statistic is conditionally asymptotically standard normal. 

Finally, it follows from the strong law of large numbers that $S$ occurs almost surely.  Consequently, the convergence in distribution to a standard normal random variable occurs almost surely.  Therefore, Polya's theorem gives that 
$$\lim_{n \rightarrow \infty} \sup_{t \in \mathbb{R}}|\hat P_{n_A, n_B}(t) - \Phi(t)| =  0$$
almost surely, which implies the result of the theorem. 

\subsection{Reproducibility}
A GitHub repository containing all relevant code and data can be found at \href{https://github.com/AnonKDD/fairness-testing}{https://github.com/AnonKDD/fairness-testing}. Also, we will open-source a Scala/Spark implementation that can be applied to large-scale data very soon!

\end{document}